\documentclass{article}
\usepackage{spconf,amsmath,graphicx}
\usepackage{enumerate}
\usepackage{amsbsy}
\usepackage{amssymb}
\usepackage{amsthm}
\usepackage{amscd}
\usepackage{subfigure}
\usepackage{color}
\usepackage{cite} 
\usepackage{dsfont}

\usepackage{balance}

\def\c1{{\textcircled{a}}}

\def\bb{{\boldsymbol{b}}}

\def\bh{{\boldsymbol{h}}}

\def\bn{{\boldsymbol{n}}}

\def\bp{{\boldsymbol{p}}}

\def\br{{\boldsymbol{r}}}

\def\bt{{\boldsymbol{t}}}

\def\bx{{\boldsymbol{x}}}
\def\by{{\boldsymbol{y}}}
\def\bz{{\boldsymbol{z}}}

\def\bC{{\boldsymbol{C}}}

\def\bH{{\boldsymbol{H}}}
\def\bI{{\boldsymbol{I}}}

\def\bW{{\boldsymbol{W}}}

\def\btau{{\boldsymbol{\tau}}}
\newcommand{\mbf}[1]{\mathbf{#1}}

\title{Deep Signal Recovery with One-bit Quantization}

\name{
	Shahin~Khobahi$^{\,\star\dagger}$, Naveed Naimipour$^{\,\dagger}$, Mojtaba~Soltanalian$^{\,\dagger}$ and Yonina~C.~Eldar$^{\,+}$
	\thanks{$^\star$ Corresponding author (e-mail: \textit{skhoba2@uic.edu}). This work was supported in part by U.S. National Science Foundation Grants CCF-1704401 and ECCS-1809225.}}
\address{$^{\dagger}$Department of Electrical and Computer Engineering,
	University of Illinois at Chicago,
	Chicago, USA\\
	$^{+}$Department of Electrical Engineering, Technion, Israel Institute of Technology, Haifa, Israel}

\begin{document}
%
\maketitle
\begin{abstract}
Machine learning, and more specifically deep learning, have shown remarkable performance in sensing, communications, and inference. In this paper, we consider the application of the deep unfolding technique in the problem of signal reconstruction from its one-bit noisy measurements. Namely, we propose a model-based machine learning method and unfold the iterations of an inference optimization algorithm into the layers of a deep neural network for one-bit signal recovery. The resulting network, which we refer to as \textit{DeepRec}, can efficiently handle the recovery of high-dimensional signals from acquired one-bit noisy measurements. The proposed method results in an improvement in accuracy and computational efficiency with respect to the original framework as shown through numerical analysis.
\end{abstract}
\section{Introduction}
\label{sec:intro}
Quantization of signals of interest is an integral part of all modern digital signal processing applications such as sensing, communication, and inference. In an ideal hardware implementation of a quantization system, a high-resolution analog-to-digital converter (ADC) with $b$-bit resolution and sampling frequency of $f_s$ samples the original analog signal and maps the obtained samples into a discrete state space of size $2^b f_s$. Generally, a large number of bits is required to obtain an accurate digital representation of the analog signal. In such a case, the quantization process has negligible impact on the performance of algorithms which were typically developed on the assumptions of infinite precision samples, and thus, the high-resolution (in terms of amplitude) quantization process can be directly modeled as an additive noise source. However, a crucial obstacle with modern ADCs is that their power consumption, manufacturing cost, and chip area grows exponentially with their resolution $b$ \cite{le2005analog, shlezinger2018hardware,kipnis2018fundamental}. 

The required high sampling data rate of ADCs used in next generataion communications systems is another obstacle that must be tackled in such applications. For instance, the promising millimeter wave (mmWave) multiple-input multiple output (MIMO) communication technology requires a very large bandwidth, and the corresponding sampling rate of the ADCs must increase accordingly. However, manufacturing ADCs with high-resolution (e.g., more than 8 bits) and high sampling rate are extremely costly and may not be available. Moreover, in other applications such as spectral sensing and cognitive radio, which require extremely high sampling rates, the cumulative cost and power consumption of using high-resolution and extremely fast ADCs are typically prohibitive and impractical. Hence, when signals across a wide frequency band are of interest, a fundamental trade-off between sampling rate, amplitude quantization precision, cost, and power consumption is encountered. An immediate solution to such challenges is to use low-resolution, and specifically \textit{one-bit}, ADCs. The use of one-bit signed measurements, and more specifically one-bit ADCs, allows for an extremely high sampling rate at a low cost and low power consumption. From a sampling viewpoint, the most extreme case of quantization is to use only one bit per sample. More precisely, one-bit sampling can be seen as a process through which we repeatedly compare the amplitude of a signal (at each sample) to some reference threshold level and use only one bit to convey whether the signal amplitude resides above or below that threshold. Due to its appealing sampling properties, the problem of recovering a signal from its one-bit measurements has attracted a great deal of interest over the past few years \cite{gianelli2016one,naveed2018onebit, li2017channel,khobahi2018onebit,liu2018massive}. Therefore, it is vital to develop algorithms that can deal with low-resolution samples for different applications.

The fields of machine learning (ML), and more particularly deep learning, are impacting various fields of engineering and have recently attracted a great deal of attention in tackling long-standing signal processing problems. The advent of low-cost specialized powerful computing resources (e.g., GPUs, and more recently TPUs) and the continually increasing amount of massive data generated by the human population and machines, in conjunction with the new optimization and learning methods, have paved the way for deep neural networks (DNNs) and machine learning-based models to prove their effectiveness in many engineering areas (see, e.g., \cite{lecun2015deep,he2015delving,deng2014deep} and the references therein). 

The main advantage of the deep learning-based model herein is that it employs several non-linear transformations to obtain an abstract representation of the underlying data. Model-based machine learning frameworks (e.g., probabilistic graphical models) incorporate prior knowledge of the system parameters into the inference process. A recent promising approach in bridging the gap between deep learning-based and model-based methods is the paradigm of \textit{deep unfolding} \cite{hershey2014deep}. Particularly, iterations of a conventional recursive algorithm, such as fast iterative soft thresholding algorithm (FISTA), projected gradient descent, and approximate message passing (AMP), can be used as a baseline to design the architecture of a deep network with trainable parameters specifically customized to the problem of interest. Such a methodology results in an improvement in accuracy, and computational efficiency of the original framework. The deep unfolding method has already shown remarkable performance improvement in a wide range of applications such as MIMO communications \cite{he2018model,samuel2018learning}, multi-channel source separation \cite{wisdom2016deep}, and sparse inverse problems \cite{gregor2010learning, borgerding2016onsager}.

In this paper, we consider the general problem of high-dimensional signal recovery from random one-bit measurements. Specifically, we propose an efficient signal recovery framework based on the deep unfolding technique that has the advantage of low-complexity and near-optimal performance compared to traditional methods. Our proposed inference framework has a wide range of applications in the areas of wireless communications, detection and estimation, and sensing.
\section{Problem Formulation}
We begin by considering a general linear signal acquisition and one-bit quantization model with time-varying thresholds, described as follows:  
\begin{align}
\label{eq:1}
&\text{Signal Model:} &&\by = \bH\bx + \bn, \\
&\text{Quantization Model:} &&\br \triangleq \text{sign}(\by - \btau),
\label{eq:2}
\end{align}
where $\btau=[\tau_1,\dots,\tau_M]^T$ denotes the vector of one-bit quantization thresholds, $\by\in\mathds{R}^M$ denotes the received signal prior to quantization, $\bH\in\mathds{R}^{M\times N}$ denotes the sensing matrix, $\bx\in\mathds{R}^{N}$ denotes the multidimensional unknown vector to be recovered, and $\bn\sim\mathcal{N}(\mbf{0},\bC)$ denotes the zero-mean Gaussian noise with a known covariance matrix $\bC=\textbf{Diag}(\sigma_1^2,\dots,\sigma_M^2)$. Furthermore, $\text{sign}(\cdot)$ denotes the signum function applied element-wise on the vector argument.

The above model covers a wide range of applications. For instance, the described model \eqref{eq:1}-\eqref{eq:2} can be used in MIMO communication systems in which $\bH$ is the channel matrix, $\bx$ is the
signal sent by the transmitter, $\bn$ is the additive Gaussian noise in the system, and the base station is equipped with one-bit ADCs, where the goal is to recover the transmitted symbols from $\br$.

\subsection{Maximum Likelihood Estimator Derivation}
Given the knowledge of the sensing matrix $\bH$, noise covariance $\bC$, and the corresponding quantization thresholds $\btau$, our goal is to recover the original (likely high-dimensional) signal $\bx$ from the one-bit random measurements $\br$. In such a scenario, each binary observation $\{r_i\}_{i=1}^{N}$ follows a Bernoulli distribution with parameter $p_i$, given by:
\begin{equation}
p_i = \text{Prob}\{ \bh_i^T\bx + n_i - \tau_i>0\} = Q\left( \frac{\tau_i - \bh_i^T\bx}{\sigma_i} \right),
\end{equation}
where $Q(x) = 1 - \Phi(x)$ with $\Phi(x)$ representing the cumulative distribution function (cdf) of a standard Gaussian distribution and $\bh_i^T$ denotes the $i$-th row of the matrix $\bH$. In particular, the probability mass function (pmf) of each binary observation can be compactly expressed as:
\begin{equation}
p(r_i) = Q\left(\frac{r_i}{{\sigma_i}}\left(\tau_i - \bh_i^T\bx\right)\right),
\end{equation}
where $r_i\in\{-1,+1\}$.
Therefore, the log-likelihood of the quantized observations $\br$ given the unknown vector $\bx$ can be expressed as:
\begin{align}
\label{eq:6}
\mathcal{L}(\bx) = p(\br|\bx) &= \text{log} \left\{ \boldsymbol{\prod}_{i=1}^{N} Q\left(\frac{r_i}{{\sigma_i}}\left(\tau_i - \bh_i^T\bx\right)\right)\right\}\\
& = \sum_{i=1}^{N}\text{log}  \left\{  \,Q\left(\frac{r_i}{{\sigma_i}}\left(\tau_i - \bh_i^T\bx\right)\right)\right\},
\end{align}
where $\text{log}\left\{\cdot\right\}$ denotes the natural logarithm. As a result, the maximum likelihood (ML) estimation of $\bx$ can be obtained as
\begin{equation}
\label{eq:7}
\hat{\bx} = \underset{\bx}{\text{argmax}}\; \mathcal{L}(\bx).
\end{equation}
Observe that the maximum likelihood estimator $\hat{\bx}$ has to satisfy the following condition:
\begin{equation}
\label{eq:8}
\nabla_{\bx}\mathcal{L}(\bx) = \mbf{0},
\end{equation}
where the gradient of the log-likelihood function with respect to the unknown vector $\bx$ can be derived as follows:
\begin{align}
\label{eq:9}
\nabla_{\bx}\mathcal{L}(\bx) = &\sum_{i=1}^{N}\left[-\frac{r_i}{\sigma_i}   \left(   \frac{ Q^{\prime}\left(\frac{r_i}{{\sigma_i}}\left(\tau_i - \bh_i^T\bx\right)\right) }{ Q\left(\frac{r_i}{{\sigma_i}}\left(\tau_i - \bh_i^T\bx\right)\right) }\right)\right]\bh_i,
\end{align}
where $Q^\prime(x) = - \frac{1}{\sqrt{2\pi}}\text{exp}\left(-x^2/2\right)$.
It can be observed from \eqref{eq:9} that the gradient of the log-likelihood function is a linear combination of the rows of the sensing matrix $\bH$. Let $\boldsymbol{\eta}\colon\mathds{R}^M \mapsto \mathds{R}^M$ be a non-linear function defined as follows:
\begin{equation}
\boldsymbol{\eta}(\bx) \triangleq \frac{Q^{\prime}(\bx)}{Q(\bx)},
\end{equation}
where the functions $Q(\cdot)$, $Q^{\prime}(\cdot)$, and the division, are applied element-wise on the vector argument $\bx$.
In addition, let $\boldsymbol\Omega = \textbf{Diag}(r_1,\dots,r_M)$ be a diagonal matrix containing the one-bit observations and $\tilde{\boldsymbol{\Omega}} = \boldsymbol{\Omega}\bC^{-\frac{1}{2}}$ be the semi-whitened version of the \emph{one-bit matrix} $\boldsymbol{\Omega}$. Then, the gradient of the likelihood function in \eqref{eq:9} can be compactly written as follows:
\begin{equation}
\label{eq:12}
\nabla_\bx\mathcal{L}(\bx) = - \bH^T\tilde{\boldsymbol{\Omega}}\,\boldsymbol{\eta}\left(\tilde{\boldsymbol{\Omega}}(\btau-\bH\bx)\right).
\end{equation}
Recall that the ML estimator $\hat{\bx}$ must satisfy the condition given in \eqref{eq:8}, i.e.,
\begin{equation}
\label{eq:13}
\nabla_\bx \mathcal{L}(\bx) = - \bH^T\tilde{\boldsymbol{\Omega}}\,\boldsymbol{\eta}\left(\tilde{\boldsymbol{\Omega}}(\btau-\bH\bx)\right) = \mbf{0}.
\end{equation}
Other than certain low-dimensional cases, finding a closed-form expression for $\hat{\bx}$ that satisfies \eqref{eq:13} is a difficult task \cite{ivrlac2007mimo,mezghani2010multiple, mo2014channel}. Therefore, we resort to iterative methods in order to find the ML estimate, i.e., to solve \eqref{eq:7}.

In this paper, the well-known gradient ascent method is employed to iteratively solve \eqref{eq:7}. Namely, given an initial point $\bx^{(0)}$, the update equation at each iteration is given by:
\begin{align}
\label{eq:14}
\bx^{(k+1)} &= \bx^{(k)} + \delta^{(k)} \nabla_\bx \mathcal{L}(\bx)\\
&= \bx^{(k)} -\delta^{(k)}\bH^T\tilde{\boldsymbol{\Omega}}\,\boldsymbol{\eta}\left(\tilde{\boldsymbol{\Omega}}(\btau-\bH\bx^{(k)})\right)
\label{eq:15},
\end{align}
where $\delta^{(k)}$ is the step size at the $k$-th iteration. The obtained maximum likelihood estimator derived from the signal model, and the corresponding optimization steps, can be unfolded into a multi-layer deep neural network, which improves the accuracy and computational effciency of the original framework.

In the next section, we \textit{unfold} the above iterations into the layers of a deep neural network where each layer denotes one iteration of the above optimization method. Interestingly, we fix the complexity budget of the inference framework (via fixing the number of layers), and apply the gradient descent method to yield the most accurate estimation of the parameter in at most $K$ iterations.
\section{Signal Recovery via Deep Unfolding}
Conventionally, first-order optimization methods, such as gradient descent algorithms, have slow convergence rate, and thus take a large number of iterations to converge to a solution. Herein, we are interested in finding a good solution under the condition that the complexity of the inference algorithm is fixed. This is important since, via unfolding the optimization algorithm, we fix the computational complexity of the inference model (a DNN with $K$ layers in such a case) and optimize the parameters of the network to find the best possible estimator with a fixed-complexity constraint. Below, we introduce \textit{\textbf{DeepRec}}, our deep learning based signal recovery framework which is designed based on the iterations of the form \eqref{eq:15}, to find the maximum likelihood estimation of the unknown parameter.\vspace{2pt}\\
\emph{---}\textbf{The \textit{DeepRec} Architecture.} The construction of DeepRec involves the unfolding of $k=1,\dots,K,$ iterations each of which are of the form \eqref{eq:15}, as the layers of a deep neural network. Particularly, each step of the gradient descent method depends on the previous signal estimate $\bx^{(k)}$, the step size $\delta^{(k)}$, the scaled one-bit matrix $\tilde{\boldsymbol{\Omega}}$, the sensing matrix $\bH$, and the threshold vector $\btau$. In addition, the form of the gradient vector \eqref{eq:12} makes it convenient and insightful to unfold the iterations onto the layers of a DNN in that each iteration of the gradient descent method is a linear combination of the system paramteres followed by a non-linearity. The $k$-th layer of DeepRec can be characterized via the following operations and variables:
\begin{align}
&\bz^{(k)} = \bW_{1k}\tilde{\boldsymbol{\Omega}}\btau-\bW_{2k}\bH\bx^{(k)} + \bb_{1k}\label{eq:16},\\
&\bp^{(k)} = \boldsymbol{\eta}\left(\bz^{(k)}\right)\label{eq:17},\\
&\bt^{(k)} = \bH^T\tilde{\boldsymbol{\Omega}}\,\bp^{(k)}\label{eq:18},\\
&\bx^{(k+1)} = f\left( \bW_{3k}
\begin{bmatrix}
\bx^{(k)}\\
\bt^{(k)}
\end{bmatrix}		 + \bb_{2k}
\right)\label{eq:19},
\end{align}
where $\bx^{(1)} = \mbf{0}$, $f(\cdot)$ denotes a non-linear activation function where in this work we consider $f(x) \triangleq \text{ReLU}(x)=\text{max}\{0,x\}$, and the goal is to optimize the DNN parameters, described as follows:
\begin{equation}
	\boldsymbol\Xi = \{\bW_{1k},\bW_{2k},\bW_{3k},\bb_{1k},\bb_{2k}\}_{k=1}^{K}.
\end{equation}

\begin{figure*}[ht]
	\begin{minipage}[b]{0.34\linewidth}
		\centering 
		\centerline{\includegraphics[width=6.4cm]{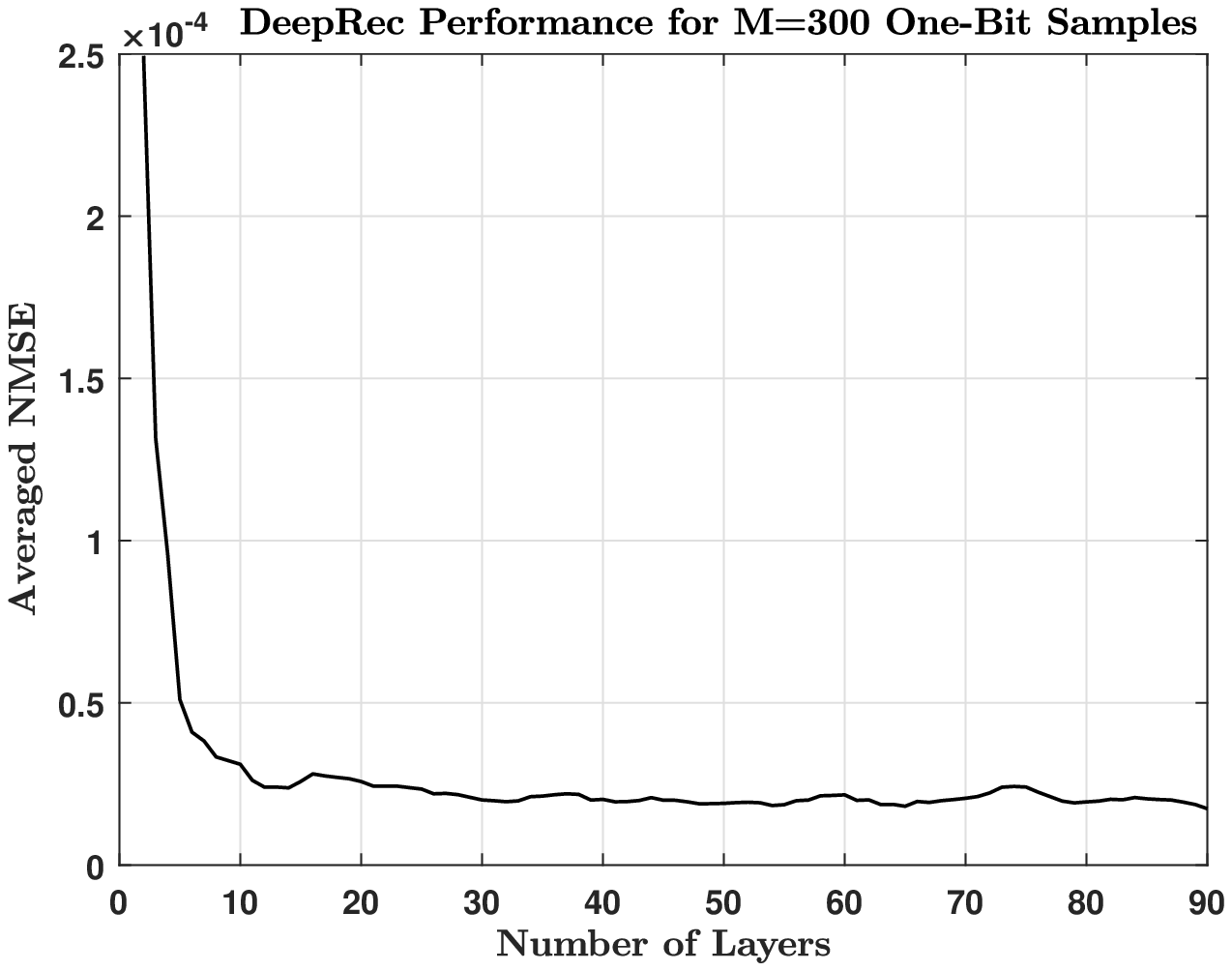}}
		\centerline{\textbf{(a)}}\medskip
	\end{minipage}
	\begin{minipage}[b]{0.34\linewidth}
		\centering
		\centerline{\includegraphics[width=6.7cm]{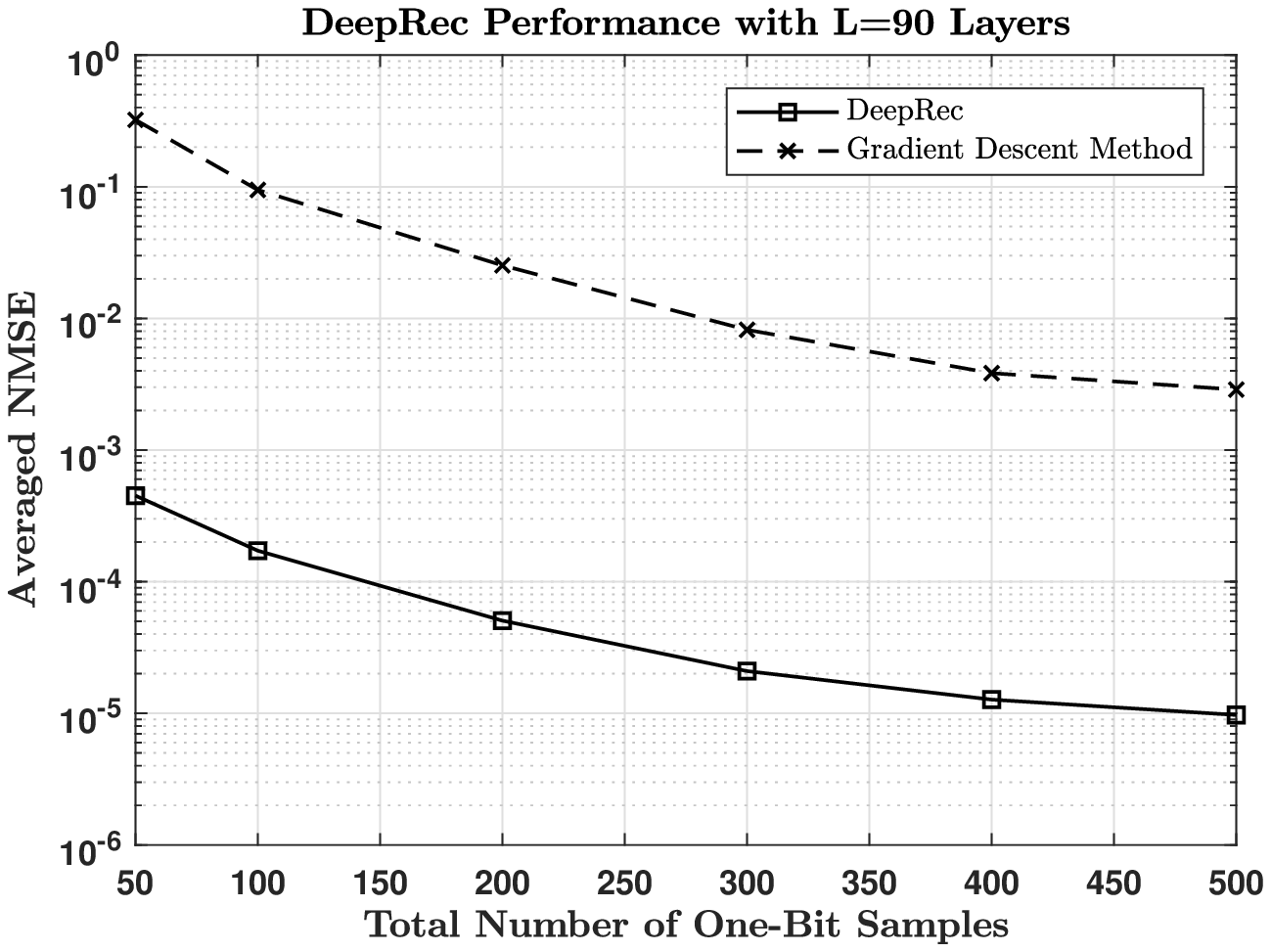}}
		\centerline{\textbf{(b)}}\medskip
	\end{minipage}
	\begin{minipage}[b]{0.34\linewidth}
		\centering
		\centerline{\includegraphics[width=6.1cm]{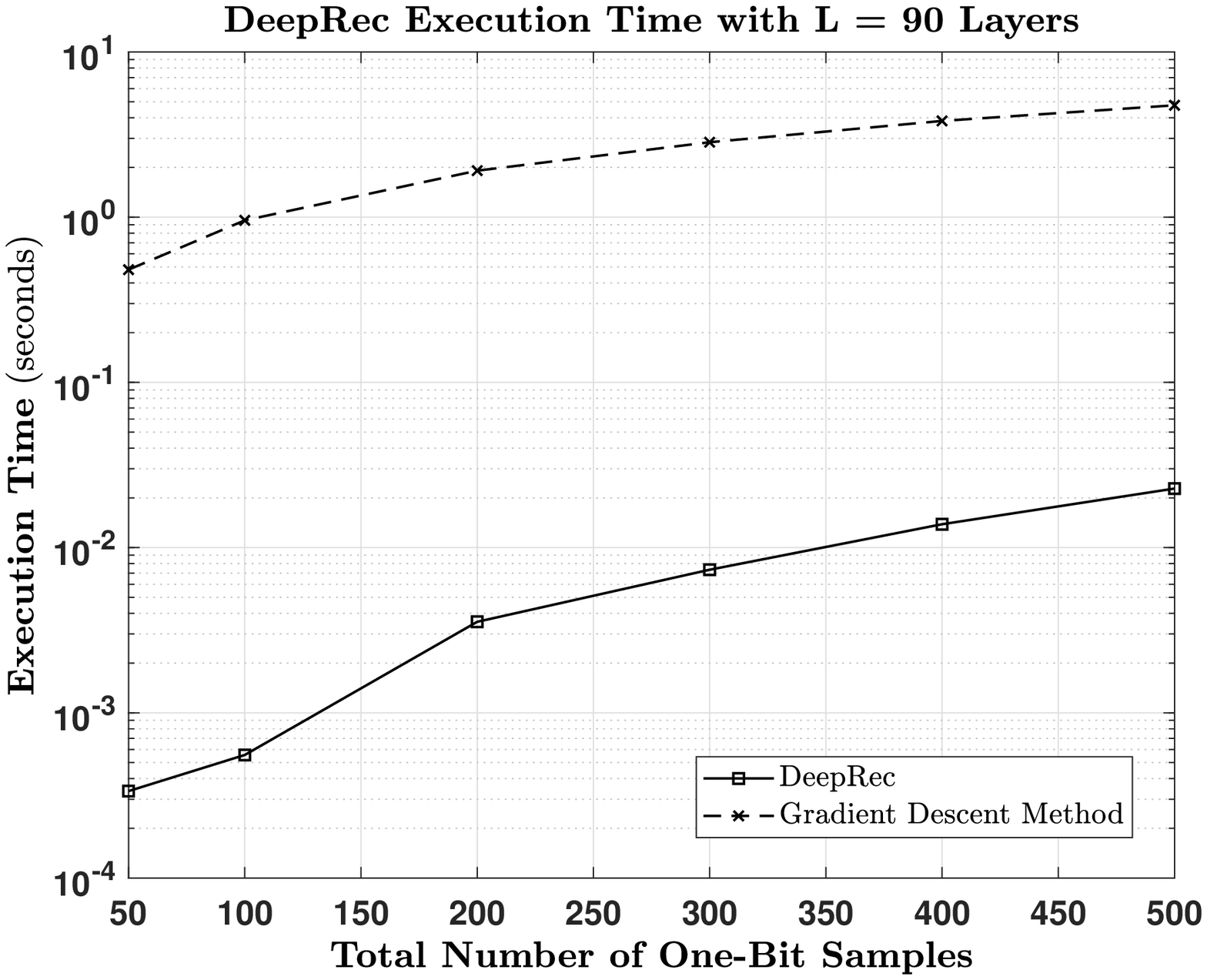}}

		\centerline{\textbf{(c)}}\medskip
	\end{minipage}
\caption{The performance of DeepRec: \textbf{(a)} demonstrates the NMSE performance of the DeepRec network for different numbers of layers $K$. \textbf{(b)} shows the performance of the proposed DeepRec architecture and the original gradient descent method of \eqref{eq:15} in terms of averaged NMSE for different numbers of one-bit samples $M$. \textbf{(c)} shows a comparison of the computational cost between the gradient descent method and the proposed DeepRec network for different numbers of one-bit samples  $M$.} 
\end{figure*}

The proposed DeepRec architecture with $L$ layers can be interpreted as a class of estimator functions $\boldsymbol\Psi_{\boldsymbol\Xi}(\br,\bH,\btau)$ parametrized by $\boldsymbol\Xi$ to estimate the unknown parameter $\bx$ given the system parameters. In order to find the best estimator function $\boldsymbol\Psi_{\boldsymbol\Xi}(\br,\bH,\btau)$ associated with our problem, we conduct a learning process via minimizing a loss function $\mathcal{R}(\bx;\boldsymbol\Psi_{\boldsymbol\Xi}(\br,\bH,\btau))$, i.e.,
\begin{equation}
\underset{\boldsymbol{\Xi}}{\text{min}}\quad \mathcal{R}(\bx;\boldsymbol\Psi_{\boldsymbol\Xi}(\br,\bH,\btau)).
\end{equation}
In this work, we employ the following least squares (LS) loss function:
\begin{equation}
\mathcal{R}\left(\bx;\boldsymbol\Psi_{\boldsymbol\Xi}(\br,\bH,\btau)\right) = ||\bx - \boldsymbol\Psi_{\boldsymbol\Xi}(\br,\bH,\btau)||_2^2,
\end{equation}
where during the training phase, we synthetically generate the system parameters $\boldsymbol\Theta = \{\bx, \br,\bH,\btau\}$ according to their statistical model.
\section{Numerical Results}
We now demonstrate the performance of the proposed DeepRec framework for the problem of one-bit signal recovery. The proposed framework was implemented using the TensorFlow library \cite{abadi2016tensorflow}, with the ADAM stochastic optimizer \cite{kingma2014adam} and an exponential decaying step size. In the learning process of the network, we employed the batch training method with a batch size of $500$ at each epoch and we performed the training for $2000$ epochs. In all of the simulations, we assumed $N=3$, i.e., $\bx\in \mathds{R}^3$, and we used the normalized mean square error (NMSE) defined as $NMSE = ||\bx-\hat\bx||_2^2/||\bx||_2^2$, for the performance metric.

The training was performed based on the data generated via the following model. Each element of the vector $\bx$ is assumed to be i.i.d and uniformly distributed, i.e., $\bx\sim\mathcal{U}(\delta_l^\bx,\delta_u^{\bx})$. The sensing matrix is assumed to be fixed and follow a Normal distribution, where we consider $\bH\sim~\mathcal{N}(\mbf{0},\bI)$.
The quantization thresholds were also generated from a uniform distribution, $\btau\sim\mathcal{U}(\delta_l^\btau,\delta_u^{\btau})$, where the lower and upper bound of the distribution is chosen in a fashion that at least covers the domain of $\bx$.
The noise is assumed to be independent from one sample to another and follows a Normal distribution, where the variance of each corresponding noise element is different, e.g., the noise covariance $\bC=~\textbf{Diag}(\sigma_1^2,\dots,\sigma_M^2)$, with $\sigma_i^2\sim\mathcal{U}(\delta_1^{\bn},\delta_M^{\bn})$. Note that we trained the network over a wide range of noise powers in order to make the DeepRec network more robust to noise.

Fig. 1(a) demonstrates the performance of the DeepRec network for different numbers of layers $K$. It can be observed that the averaged NMSE decreases dramatically as the number of layers increases. Such a result is also expected as each layer corresponds to one iteration of originial optimization algorithm. Thus, as the number of layers increases, the output of the network will converge to a better estimation as well.

Fig. 1(b) demonstrates the performance of the proposed DeepRec architecture and the original Gradient Descent method of \eqref{eq:15} in terms of averaged NMSE for different numbers of one-bit samples $M$. In this simulation, we implemented the DeepRec network with $K=90$ layers. It can be clearly seen from Fig. 1(b) that the proposed deep recovery architecture (DeepRec) significantly outperforms the original optimization method in terms of accuracy and provides a considerably better estimation than that of the gradient descent method for the same number of iterations/layers. As a fair comparison, we also assumed a fixed-step size of $\delta=0.01$ for the gradient descent method.

Fig. 1(c) shows a comparison of the computational cost (machine runtime) between the gradient descent method and the proposed DeepRec network for different numbers of one-bit samples $M$. It can be seen that our proposed method (DeepRec) has a significantly lower computational cost than that of the original optimization algorithm for our problem. Hence, making the DeepRec a good candidate for real-time signal processing or big data applications (the results were obtained on a standard PC with a quad-core 2.30GHz CPU and 4 GB memory).
\section{Conclusion}
We have considered the application of model-based machine learning, and specifically the deep unfolding technique, in the problem of recovering a high-dimensional signal from its one-bit quantized noisy measurements via random thresholding. We proposed a novel deep architecture, which we refer to as \textit{DeepRec}, that was able to accurately perform the task of one-bit signal recovery. Our numerical results show that the proposed DeepRec network significantly improves the performance of traditional optimization methods both in terms of accuracy and efficiency. 
\bibliographystyle{IEEEbib}
\bibliography{strings}
\end{document}